%% file: DZ3.tex
\documentclass{aa}

\newcommand{\pv}{\ensuremath{P_V}}
\newcommand{\nv}{\ensuremath{N_V}}

\newcommand{\bz}{\ensuremath{\langle B_z \rangle}}
\newcommand{\nnz}{\ensuremath{\langle N_z \rangle}}

\newcommand{\bs}{\ensuremath{\langle \vert B \vert \rangle}}

\newcommand{\snr}{\ensuremath{S/N}}

\newcommand{\teff}{\ensuremath{T_{\rm eff}}}

\usepackage{graphicx}
\usepackage{txfonts}
\bibpunct{(}{)}{;}{a}{}{,}
\begin{document}

   \title{Discovery of weak magnetic fields in four DZ white dwarfs\\ in the local 20\,pc volume}
\subtitle{Implications for the frequency of magnetic fields with cooling age}

   \author{S. Bagnulo \inst{1}
          \and
          J.D. Landstreet\inst{1,2}
          }

   \institute{Armagh Observatory and Planetarium, College Hill, Armagh BT61 9DG, UK \\
              \email{stefano.bagnulo@armagh.ac.uk; john.landstreet@armagh.ac.uk}
         \and
             University of Western Ontario, London, Ontario N6A 3K7, Canada.
             \email{jlandstr@uwo.ca}
             }

   \date{Received June 11, 2019; accepted August 16, 2019}

  \abstract{
  We report the discovery of weak magnetic fields in three white dwarfs within the local 20\,pc volume (WD\,0816$-$310, WD\,1009$-$184, and WD\,1532$+$129), and we confirm the magnetic nature of a fourth star (WD\,2138$-$332) in which we had previously detected a field at a 3\,$\sigma$ level. The spectra of all these white dwarfs are characterised by the presence of metal lines and lack of H and He lines, that is, they belong to the spectral class DZ. The polarisation signal of the Ca\,{\sc ii} H+K lines of WD\,1009$-$184 is particularly spectacular, with an amplitude of 20\,\% that is due to the presence of a magnetic field with an average line-of-sight component of 40\,kG. We have thus established that at least 40\,\% of the known DZ white dwarfs with an He-rich atmosphere contained in the 20\,pc volume have a magnetic field, while further observations are needed to establish whether the remaining DZ white dwarfs in the same volume are magnetic or not. \\
  Metal lines in the spectra of DZ white dwarfs are thought to have originated by accretion from rocky debris, and it might be argued that a link exists between metal accretion and higher occurrence of magnetism. However, we are not able to distinguish whether the magnetic field and the presence of a polluted atmosphere have a common origin, or if it is the presence of metal lines that allows us to detect a higher frequency of magnetic fields in cool white dwarfs, which would otherwise have featureless spectra. \\
  We argue that the new highly sensitive longitudinal field measurements that we have made in recent years are consistent with the idea that the magnetic field appears more frequently in older than in younger white dwarfs.
  }

   \titlerunning{Discovery of weak magnetic fields in four DZ white dwarfs in the local 20\,pc volume}

   \maketitle
%

\section{Introduction}\label{Sect_Intro}
Magnetic fields are present in a non-negligible fraction of white dwarfs (WDs). In different stars, the observed field strength spans more than five orders of magnitude, from a few kiloGauss to almost a gigaGauss. These fields have a globally organised structure that is relatively stable with time, at least on a timescale of the order of decades. The observed variability of the magnetic field is due to the fact that the magnetic geometry is often not symmetric about the rotation axis, so that the observer sees a magnetic configuration that changes as the star rotates. Typical rotation periods are between minutes and weeks \citep[e.g.][]{Brietal13}, although there are cases of extremely long rotation periods, such as those of WDs GD\,229, GD\,240$-$72, and Grw\,+70\degr\,8247, which show polarimetric variability on a timescale of the order of ten years or more, and must have a rotation period longer than one century \citep{BerPii99,BagLan19}. 

The magnetic field does not seem to be maintained by any current dynamo action, but appears to be the remnant of a field generated during earlier stages of stellar evolution that has survived a very slow Ohmic decay \citep[$\sim 10^{10}$\,yr,][]{Wenetal87}. The origin of the magnetic field is still very uncertain. It may have been generated as a result of dynamo action in the convective core during an earlier stage of stellar evolution, or it may be the result of the intense interaction during a binary system merger \citep{Touetal08}. Multiple evolution paths may lead to presence or absence of a global magnetic field in a white dwarf. 

Most studies of magnetic WDs (MWDs) have been based on target lists that are biased by certain factors (e.g. apparent magnitude of the star), and with techniques having a wide range of sensitivity. Consequently, field detection has also been biased by various factors, mainly favouring, for obvious reasons, the strongest fields and the brighter WDs. Field strength is an especially strong bias for the WDs that have been discovered to be magnetic by observation of Zeeman splitting detected in low-resolution spectra that have a low signal-to-noise ratio (\snr), such as those of the Sloan Digital Sky Survey (SDSS) \citep[see the discussion in Sect.~2.6 of][]{BagLan18}. In order to characterise the global qualities of the magnetic fields of WDs and identify their relationships to other stellar parameters, a sample needs to be surveyed that is defined by criteria that are independent of magnitude and spectral type, such as the sample that currently lies within 20\,pc of the Sun \citep{Holbetal16,Holletal18}. This volume of space contains about 150 WDs; if needed in future, statistics can be improved by simply increasing the volume size. 
At the present time, magnetic data about the WDs in the 20\,pc volume sample are still incomplete. While some of the WDs within the 20\,pc volume have been closely examined for magnetic fields over the years and a wide variety of fields have been discovered, the available data for many of the WDs were only sufficient to identify fields above about 1\,MG until a few years ago. Because many MWDs have fields $\ll 1$\,MG, we know that neither low-resolution spectroscopy nor even high-resolution spectroscopy are practically able to strongly constrain the possible presence or absence of a magnetic field. 

In response, we have been actively observing WDs by a variety of methods in order to obtain more complete statistics of the MWDs in the 20\,pc volume. In the course of our survey we have already identified a few new MWDs \citep[see e.g.][and references therein]{LanBag19b}. Here we focus on the subsample of WDs that have a spectrum without H or He lines, but exhibit metal lines, that is, WDs belonging to spectral class DZ. Most of these stars have a pure He atmosphere, but He lines are not seen because of their low temperature ($\teff \la 12\,000$\,K). H Balmer lines would also not be visible in a hydrogen-dominated atmosphere for $\teff \la 5\,000\,K$. According to Table~2 of \citet{Giaetal12} and Table~4 of \citet{Subetal17}, all known DZ stars within 20\,pc have a He-rich atmosphere, except for WD\,0552$-$041.

Metals present in the atmosphere of WDs are thought to originate from rocky circumstellar debris that has survived the post-main sequence evolution of the star \citep{Jura03}. Metal lines also appear in some WDs with H-dominated atmosphere, which have spectra with Balmer lines. In this case, the spectra are classified as DAZ.

The DZ WDs are particularly important in the study of magnetic fields in cool WDs with He-dominated atmospheres. Several types of cool WDs exist in which the outer layers are He dominated, including DC, DQ, DZ, and DZA stars. Only the DZ and DZA WDs show atomic lines in their spectra. DZ stars show no other lines except for metal lines. DZA stars also exhibit a very weak H$\alpha$ line; opacity is still mainly due to He, and H appears just as a trace element. In summary, because of their atomic spectral lines, the DZ/DZA stars are the only cool WDs with He-dominated atmospheres in which weak fields can be detected with spectroscopy or spectropolarimetry using the Zeeman effect. 

The incidence of magnetic fields in stars with metal lines has been discussed in previous publications, and has been proposed to be higher in these types of star than in WDs of other spectral types, particularly DA stars. \citet{Holletal17} found that more than 10\,\% of a sample of 231 cool ($\teff \la 9\,000$\,K) DZ WDs observed in the framework of the SDSS exhibit Zeeman splitting from a strong ($\ga 0.5$\,MG) magnetic field. This result agrees with the previous finding by \citet{Holletal15} that $13 \pm 4$\,\% of DZ stars are (strongly) magnetic. Regarding WDs with a H-rich atmosphere, \citet{Kawetal19} have suggested that DAZ stars show an incidence of magnetic field that is substantially higher than in WDs of other spectral types, claiming that $\sim 50$\,\% of DAZ WDs with $\teff \la 6\,000$\,K are magnetic. \citet{KawVen14} have explicitly suggested a link between a crowded planetary system and magnetic field generation.

In this paper we report the discovery of new MWDs, all belonging to the spectral class DZ and residing within the local 20 pc volume. We also make a preliminary estimate of the incidence of magnetic fields in cool WDs with an atmosphere that is predominantly composed of He, and with metal lines. 

\section{Instruments and instrument settings}
Our observations were obtained with the FORS2 instrument \citep{Appetal98} of the ESO Very Large Telescope (VLT) and with the ISIS instrument of the William Herschel Telescope. Our targets were observed in spectropolarimetric mode to measure circular polarisation (Stokes $V/I$) as well as the unpolarised  spectrum (intensity, Stokes $I$). Both instruments are equipped with conceptually similar polarimetric modules consisting of a retarder waveplate that may be rotated at set position angles, and a beam-splitting device (a Wollaston prism in the case of FORS2 and a Savart plate in case of ISIS) that splits the incoming radiation into two beams that are linearly polarised in orthogonal directions. With both instruments we applied the beam-swapping technique, which consists of observing with the retarder waveplate at two position angles separated by 90\degr\ (typically performing one or two cycles $-45\degr$, $+45\degr$, $+45\degr$, $-45\degr$). This technique allows the minimisation of the instrumental polarisation \citep[see e.g.][]{Bagetal09}. 

When we observed with FORS2, we used grism 1200B, which covers the spectral range 3660--5100\,\AA\ with a 1\arcsec\ slit width for a spectral resolution of $\sim 1400$. One of our targets, star WD\,2138$-332$, has previously been observed with the same instrument and same
setting in 2014 \citep[see][]{BagLan18}. To reduce overhead time and readout noise, the readout mode was set to $2 \times 2$ binning.

When we used ISIS, we observed with both the blue and the red arms simultaneously by inserting a 5300\,\AA\ dichroic mirror in the optical beam. In the blue arm we used grating R600B centred at 4400\,\AA, which approximately covers the spectral range  3650--5150\,\AA; in the red arm we used grating R1200R with order separating filter GG\,495, centred at 6270\,\AA, which covers the spectral range 5870--6670\,\AA. The setting in the red arm was slightly different than the one adopted in previous observing runs (for the observations presented by \citeauthor{BagLan18} \citeyear{BagLan18} we had set the central wavelength at 6500\,\AA) because we wished to attempt to detect the Na {\sc i} doublet at 5889--95\,\AA. The slit width was set to 1\arcsec\ for a spectral resolution of $\sim 2600$ in the blue arm around H$\beta$ and $\sim 8600$ in the red arm around H$\alpha$. The readout mode was set to $2 \times 2$ binning.

Observations were reduced as described by \citet{BagLan18}. The magnetic field values were
measured as explained in the following section.

\section{Measuring the magnetic field}\label{Sect_Field}
The mean longitudinal magnetic field \bz\ (the line-of-sight component of the surface magnetic field, averaged over the visible hemisphere of the star) and the mean field modulus \bs\ (the total field strength $|B|$, averaged over the visible stellar hemisphere) may be estimated from the stellar polarised spectra using  methods that have been discussed in previous work \citep[see][and references therein]{LanBag19b}. In particular, for the mean longitudinal magnetic field we applied a correlation method that relies on the weak-field approximation. In this regime, the right and left circularly polarised line components are only slightly shifted relative to one another, and the local fractional circular polarisation $V/I$ is proportional to the local slope of the $I$ profile through the proportionality 
\begin{equation}
\frac{V(\lambda)}{I(\lambda)} = - C_z\, g_{\rm eff} \frac{1}{I(\lambda)}\  
                            \frac{{\rm d}I(\lambda)}{{\rm d} \lambda} \bz 
\label{Eq_Bz}
,\end{equation}
where $C_z$ is a constant $= -4.67\,10^{-13}$\,\AA$^{-1}\,{\rm G}^{-1}$, $g_{\rm eff}$ is the effective Land\'e factor, and \bz, the mean longitudinal magnetic field, is measured in G. The numerous instances of this equation through the line profile are treated as a least-squares problem to be solved for \bz,\ as discussed for instance by \citet{Bagetal02}. 

As a quality check, when observations are obtained with at least two pairs of exposures, null profiles can be calculated and inspected \citep{Donetal97,Bagetal09}. They are essentially the difference between the reduced Stokes $V$ profiles obtained from independent pairs of measurements. Null profiles may be considered as an experimental estimate of the noise. Cosmic rays and instrument flexures are likely to cause the null profiles to depart from zero (e.g. to produce a spike in the proximity of a spectral line); the presence of a feature in the same spectral region both in the null and in the reduced Stokes $V$ profile is likely to be of spurious origin \citep[conversely, it is still possible that a feature seen in the Stokes $V$ profile is of spurious origin even if it does not appear in the null profile, see][]{Bagetal13}. Equation~(\ref{Eq_Bz}) may be applied to the null profiles to calculate the so-called null field \nnz, and verify that it is zero within measurement uncertainty. As extensively discussed by \citet{Bagetal12} and \citet{Bagetal13}, null field values have a statistical significance. On an individual star, measuring a null field value consistent with zero does not guarantee that the corresponding \bz\ measurement is free from spurious effects (in the same way as a flat null profile does not guarantee that a spike seen in Stokes $V$ is real). The best quality check consists of verifying that the \nnz\ values normalised by their uncertainties are distributed as a Gaussian centred about zero and with $\sigma=1$. In the case of a small number of measurements, as in the present case, we only checked that no null field measurements exceeded $2-3\,\sigma$ in absolute value.

We recall that for geometrical reasons, it is possible that a magnetic field escapes detection if at the time of the observations its line-of-sight component, averaged over the visible stellar disc, is zero. This could be the case, for instance, for a dipole field seen when its axis is perpendicular to the line of sight. For this reason, a single null \bz\ measurement does not guarantee that a star is not magnetic. In addition, spurious effects may degrade circular polarisation measurements \citep[e.g.][]{Bagetal12,Bagetal13}. Therefore it is always a good idea to confirm the magnetic nature of a star with more than one measurement, in particular when field measurements are at low level of significance. 

None of the stars discussed in this papers shows evidence of Zeeman splitting. By studying the line cores of the metal lines, we are able to set a rough upper limit on the mean field modulus \bs. Upper limits on \bs\ depend on spectral resolution, \snr, and on which metal lines are present in the spectra. We discuss this for the individual cases.

\section{Results}
\input{Table_Stars}
The DZ stars in which we have detected a magnetic field are WD\,0816$-$310, WD\,1009$-$184, WD\,1532$+$129, and WD\,2138$-$332. Their general characteristics are given in Table~\ref{Table_DZ}, and the log of our observations and field measurements are given in Table~\ref{Tab_Log}. In this section we discuss them individually (Sects.~\ref{Sect_WD0816} -- \ref{Sect_WD2138}). For the purposes of our discussion, we also briefly summarise the results of the observations of four additional DZ and DZA WDs (see Sect.~\ref{Sect_Others}).
\input{Table_Log.tex}

\subsection{WD\,0816$-$310}\label{Sect_WD0816}
\begin{figure}
\includegraphics*[width=9.3cm,trim=0.75cm 6.5cm 0cm 3cm,clip]{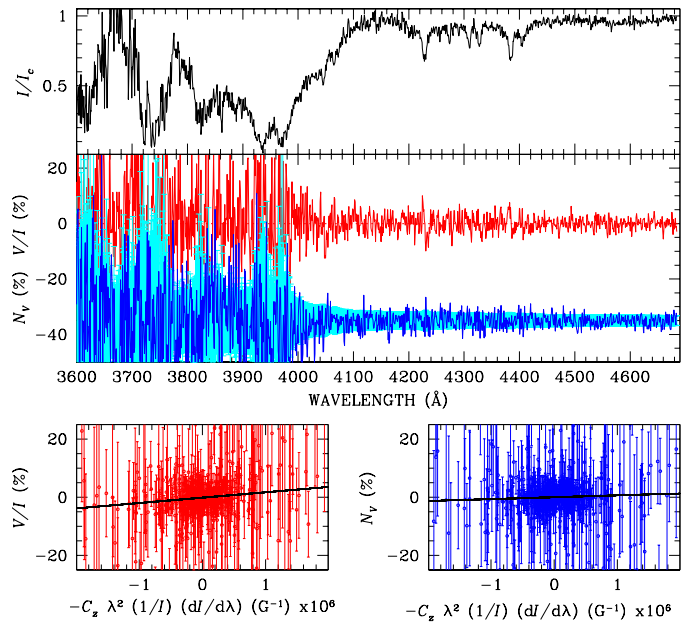}\\
\caption{\label{Fig_WD0816_ISIS} 
  Polarised spectra of WD\,0816$-$310 obtained with the blue arm of ISIS. In the upper panel, the black solid line represents the intensity profile normalised to a pseudo-continuum. The red solid line is the $V/I$ profile (in percent) and the blue solid line is the null profile (offset from zero for display purposes). Error bars are plotted on the null profiles and appear as a light blue background. The bottom panels show the best fit obtained by minimising the expression of Eq.~(\ref{Eq_Bz}) using the $V/I$ profiles (left panel) and the \nv\  profiles (right panel).
  }
\end{figure}
\begin{figure}
\includegraphics*[width=9.3cm,trim=0.75cm 6.5cm 0cm 3cm,clip]{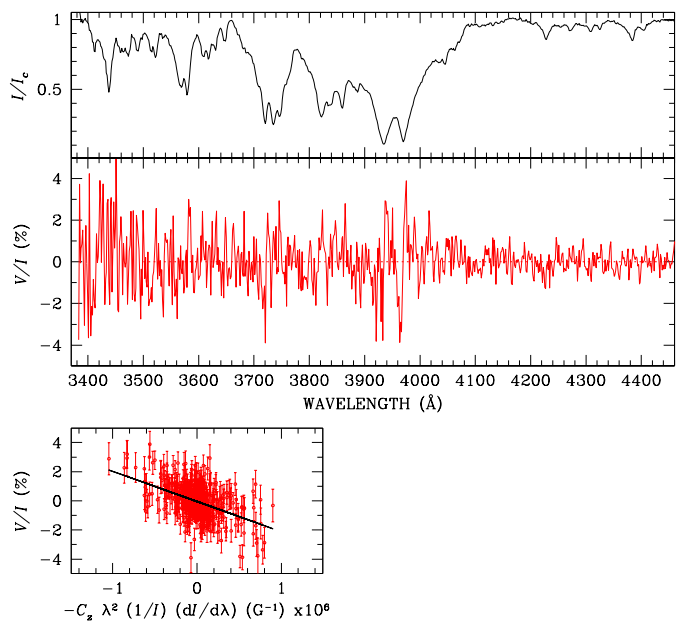}\\
\caption{\label{Fig_WD0816_FORS} 
  Polarised spectra of WD\,0816$-$310 obtained with FORS1 using grism 600B. The figure is similar to
  Fig.~\ref{Fig_WD0816_ISIS}, except that the observations were obtained with only two retarder
  waveplate positions, therefore null spectra are not available.
  }
\end{figure}

WD\,0816$-$310 = SCR\,J0818-3110 was spectroscopically confirmed to be a WD of spectral class DZ by \citet{Subetal08}. \citet{Beretal14} found no evidence of IR excess in Spitzer data.

We observed the star with ISIS during our April 2019 run. Our spectrum is extremely noisy, with a peak \snr\ $\sim 65$ per \AA\ (see the upper panel of Fig.~\ref{Fig_WD0816_ISIS}). However, the spectrum recorded in the blue arm clearly exhibits the Fe\,{\sc i} lines at 3824, 3860\,\AA\ and those between 4250 and 4400\,\AA; the Mg\,{\sc i} lines at 3832-3838\,\AA, the Ca\,{\sc ii} H+K lines at 3933 and 3968\,\AA, and the Ca\,{\sc i} line at 4227\,\AA.  The spectrum recorded in the red arm is virtually featureless. The Mg\,{\sc i} and Ca\,{\sc ii} H+K lines are so deep that their cores are sampled with an \snr\ much lower than the continuum (down to $\sim 10$ per \AA), hence the measurements of the circular polarisation of the spectral lines are affected by large errors.
In the two lower panels of Fig.~\ref{Fig_WD0816_ISIS} we show the correlation between $V/I$ and the coefficient of \bz\ (left panel) and the correlation between the null spectrum \pv\ and the same coefficient (right panel). Solving for \bz,\ we find a value of $\bz \approx 19 \pm 5$\,kG, while solving for \nnz,\ we obtain a null detection: $6.5 \pm 4.6$\,kG. In conclusion, while it is not possible to identify a clear $V$ signature associated with any spectral lines, a correlation between $V$ and d$I$/d$\lambda$ reveals a magnetic field at a $3-4\,\sigma$ level. 

We found FORS1 spectropolarimetric archive data obtained in 2007 (to the best of our knowledge, they were never published) that confirm that the star is magnetic. From these archive data, shown in Fig.~\ref{Fig_WD0816_FORS}, we have measured $\bz=-21 \pm 2$\,kG, a detection with a $10\,\sigma$ significance, which also clearly shows that the star is magnetically variable.

Our ability to detect the Zeeman splitting in Stokes $I$ that should be present in the numerous sharp absorption lines of Fe~{\sc i} in the spectral region around the Ca~{\sc ii} H\&K lines is limited by the resolving power (and \snr) of our observations. The FORS1 observations, with a resolution of about 3\,\AA, could only reveal Zeeman splitting in which the $\pi - \sigma$ separation is larger than about 3\,\AA. This splitting is produced by a magnetic field of approximately 400\,kG. Similarly, the higher resolution of the ISIS blue spectra, about 1.5\,\AA, was able to uncover Zeeman splitting produced by a field of about 200\,kG. In WD\,0816-310, the Ca~{\sc i} 4226 line and the Fe~{\sc i} 4383 lines appear in our ISIS spectrum to have quite square cores, which may show the presence of a mean field modulus \bs\ of the order of 250--300\,kG. Unfortunately, we cannot confirm this possibility using the numerous strong Fe~{\sc i} lines shortward of the Ca H\&K lines because the \snr\ ratio is too low in this region. Conservatively, then, we can set an upper limit to \bs\ of about 300\,kG.

\subsection{WD\,1009$-$184}\label{Sect_WD1009}
WD\,1009$-$184 = WT\,1759 is a cool WD of spectral type DZ, reflecting a bulk atmospheric composition of nearly pure He, slightly polluted with Ca, as revealed by the Ca\,{\sc ii} H \& K lines and several weaker nearby features \citep{Subetal09}.
WD\,1009$-$184 is a member of a wide visual binary star system. It is the common proper motion companion of the K7V star BD$-17$\,3088, with 400\arcsec\ angular separation \citep{Holbetal08,Tooetal17}. With this angular separation, the two stars are separated in space by several thousand AU, and it is very unlikely that they have ever interacted significantly, other than via the gravitational forces governing their mutual orbits. As for WD\,0816$-$310, \citet{Beretal14} did not find evidence of IR excess in the spectral energy distribution. 

\begin{figure}
\includegraphics*[width=9.3cm,trim=0.75cm 6.5cm 0cm 3cm,clip]{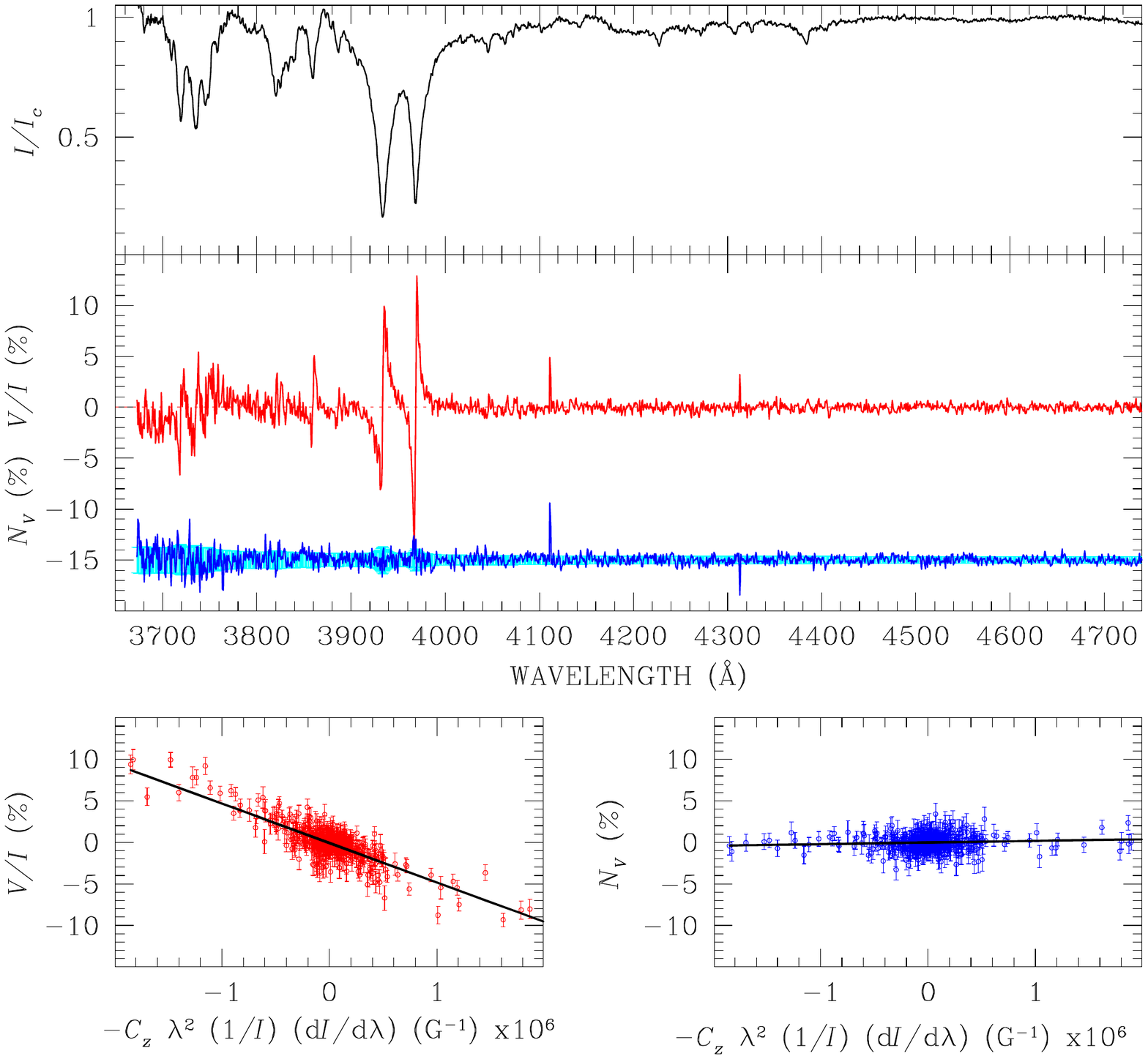}\\
\vskip 0.5cm

\includegraphics*[width=9.3cm,trim=0.75cm 6.5cm 0cm 3cm,clip]{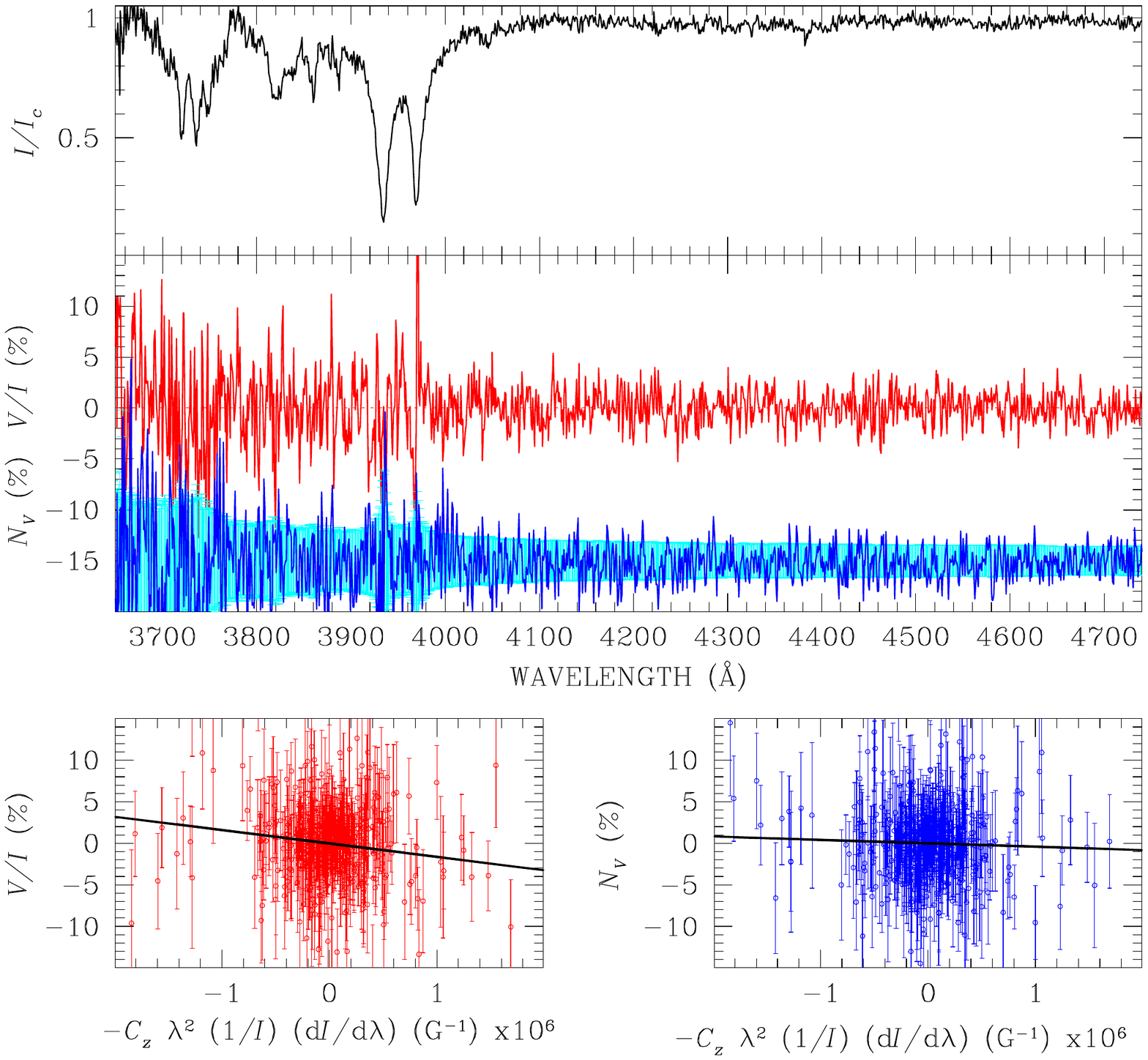}
\caption{\label{Fig_WD1009} 
{\it Top panels:} Polarised spectrum of WD\,1009$-$184 obtained in December 2018 with the FORS2 instrument; the meaning of the panels is the same as in Fig.~\ref{Fig_WD0816_ISIS} for star WD\,0816$-$310. {\it Bottom panels:} Same as above, but with data obtained in April 2019 with the blue arm of the ISIS instrument.}
\end{figure}
WD\,1009$-$184 was observed with FORS2 on December 25, 2018, and with ISIS on April 19, 2019. The FORS2 spectrum has a peak \snr\ $\sim 370$ per \AA. None of the spectral lines shows any sign of Zeeman splitting in the Stokes $I$ spectrum, so we are not able to estimate \bs. However, a strong circular polarisation signal is present in the H and K Ca lines, with a peak-to-peak amplitude of $\sim 20$\,\%; corresponding weaker signals are seen in most of the shallower spectral features. Because the observed $I$ spectrum is hardly different from what it would be in the absence of a field, the correlation method of measurement of \bz\ discussed in Sect.~\ref{Sect_Field} may be used to measure the mean longitudinal field. Using the wavelength interval $3650-4100$\,\AA, for \bz\ we find a value of $\bz = -47 \pm 1$\,kG. With a polarisation signal significant at the $30\,\sigma$ level, there is certainly no need for further observations to confirm field detection. Nevertheless, we re-observed this WD with ISIS to check for variability, and also to assess the performance of ISIS on a faint magnetic DZ star. With ISIS we used the same integration time as  we adopted for our FORS2 observations. Because the collecting area of the WHT is 3.6 times smaller than that of the VLT, we would expect  an \snr\ per \AA\ that is about twice lower. Because of much worse seeing and high background level (observations were performed during a night of full Moon), we reached an \snr\ of only 80 per \AA\ (4.5 times lower than with FORS2), and our field measurement uncertainty was four times larger than that obtained with FORS2. 

Thanks to its two arms, the spectral coverage of ISIS is roughly twice as large as that of FORS2, and its spectral resolution is also higher. However, for this star, larger spectra coverage did not lead to any advantage because the spectrum recorded in the red arm is featureless. We detected a field using the metal lines observed with the blue arm, although at just $4\,\sigma$ significance level: $\bz = -17\pm 4.5$\,kG; the null field was $\nnz =-4 \pm 4.5$\,kG. The magnetic field of the star is therefore definitely variable, and it would be an ideal target for a monitoring run for modelling purposes. 

Detectable Zeeman splitting has been searched for in the spectral lines in the blue and near-UV. In the case of WD\,1009$-$184, with a field \bz\ that can approach 50\,kG in strength, we expect a mean field modulus \bs\ to be at least about 150\,kG, which is not far below the field strength that would visibly change the Stokes $I$ (flux) line core profiles from roughly Gaussian or even pointed shapes into rather square profiles, which, with a large enough field,  would show incipient Zeeman splitting. A few of the Fe\,{\sc i} lines near 4350\,\AA\ should show significant profile changes even for fields of perhaps 200\,kG. However, in our spectra of WD\,1009$-$184 these lines are so weak that any Zeeman structure is lost in the noise. The strongest limit on \bs\ in this star is obtained from the core of the Ca\,H line in the ISIS spectrum, with a total core width of about 5\,\AA. This core might be broadened by a field of the order of about 300\,kG. At present, we can therefore only derive an upper limit to \bs\ of roughly 300\,kG. 

\begin{figure}
\includegraphics*[width=9.3cm,trim=0.75cm 6.5cm 0cm 3cm,clip]{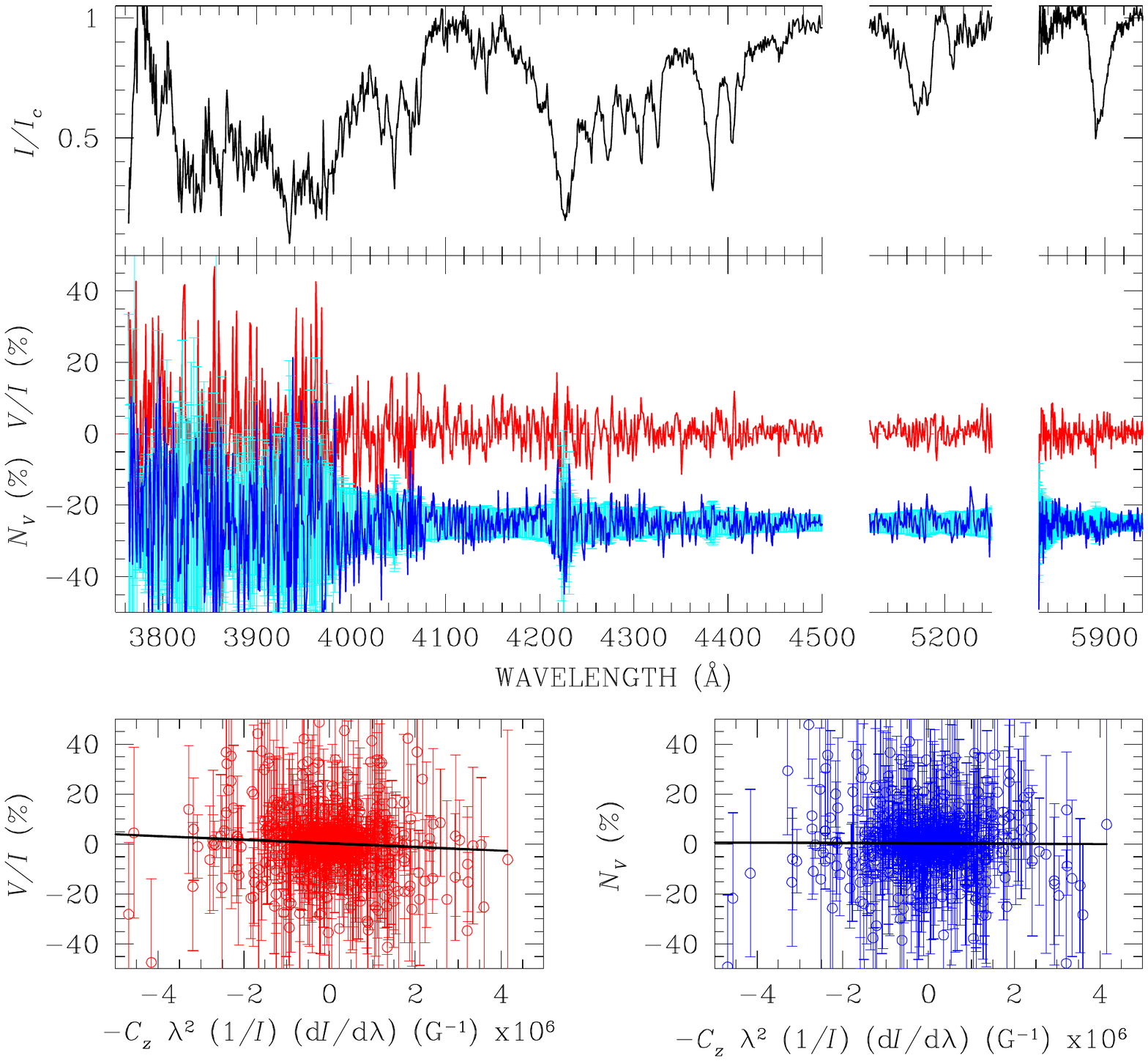}
\vskip 0.5cm

\includegraphics*[width=9.3cm,trim=0.75cm 6.5cm 0cm 3cm,clip]{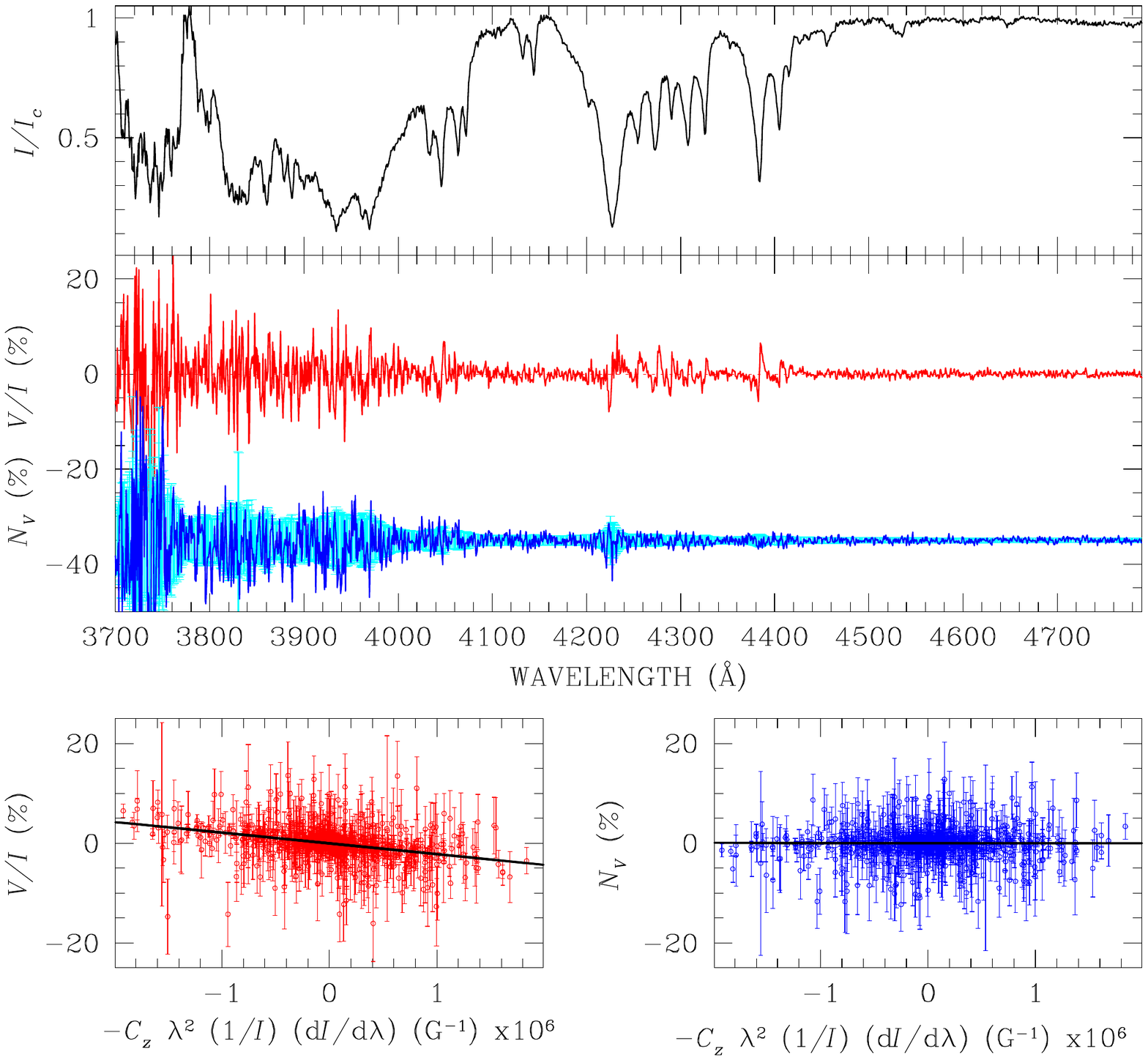}\\
\caption{\label{Fig_WD1532} 
  Polarised spectra of WD\,1532$+$129 obtained in April 2019 with the ISIS instrument
  (top panels) and in June 2019 with the FORS2 instrument (bottom panels). We show spectral ranges in different wavelengths. Both spectra were normalised to the pseudo-continuum.
  The panels are organised as for WD\,0816$-$310 in Fig.~\ref{Fig_WD0816_ISIS}.
}
\end{figure}

\subsection{WD\,1532$+$129}\label{Sect_WD1532}
WD\,1532$+$129 = G\,137-24 is the coolest star of our sample \citep[\teff = 5\,430\,K,][]{Subetal17}. It was classified as a DZ WD by \citet{Kawetal04}, who originally estimated a temperature of 7\,500\,K, noting, however, that the photometry suggested the lower value of 6\,000\,K. Strong Ca\,{\sc ii} absorption lines at 3933, 3968 and 4226\,\AA\ are clearly evident in its spectrum.  Mg\,{\sc i} absorption lines are also seen at 3829, 3832, and 3838\,\AA\ (blended), as well as the Mg\,{\sc i} triplet at 5167, 5173, and 5184\,\AA\ (also blended). Several weak Fe\,{\sc i} lines between 4000 and 4500\,\AA\ and the sodium D doublet at 5890--5896\,\AA\ (blended) are visible. \citet{Kawetal04} also reported that the spectrum of the star closely resembles that of WD\,1328$+$307, a DZ star discovered to have a magnetic field by \citet{Dufetal06}, who identified several Zeeman split lines that are visible (because of the strength of the field, with $\bs \approx 650$\,kG) even with low-resolution spectroscopy of sufficiently high \snr\ ratio.

In March 2019, with FORS2 and using grism 1200B, we obtained a spectrum with a peak $\snr \sim 250$ per \AA. Using the correlation technique of Sect.~\ref{Sect_Field}, we obtained a $4\,\sigma$ detection: $\bz = -4.3 \pm 1.0$\,kG. From the null spectrum we obtained a non-detection: $\nnz = -1.2 \pm 0.9$\,kG. In April 2019, we re-observed the star with ISIS, reaching only a
peak \snr\ of 60 per \AA\ in the blue arm, and 85 per \AA\ in the red arm, where we could detect the Na\,{\sc i} line at 5880\,\AA. A field of similar strength was detected both in the blue and in the red arm. In the blue arm spectrum we measured $\bz = -7.7 \pm 2.6$\,kG, and from the red arm spectrum we measured $\bz = -9.6 \pm 2.3$\,kG. Combining the data obtained in the two arms, we obtained $\bz=-8.5 \pm 1.9$\,G, again a 4\,$\sigma$ detection, and $\nnz = 0.2 \pm 1.8$\,kG. Finally, in June 2019, we observed
the star again with FORS2, and we measured $\bz = -21.5 \pm 0.9$\,kG  and $\nnz = -0.2 \pm 0.9$\,kG. Figure~\ref{Fig_WD1532} shows the spectra obtained with ISIS in April and with FORS2 in June 2019.
Our three field detections, in particular the one at a $\sigma$ significance level $> 20$ obtained with FORS2 in June 2019, definitely demonstrate that the star is magnetic, with one of the weakest fields ever measured in a WD.

WD\,1532$+$129 has a rich spectrum of Fe~{\sc i} lines that show no obvious Zeeman splitting. From the core widths of the narrower of these lines in our ISIS spectrum, about 5\,\AA, we derive an upper limit to \bs\ of about 300\,kG.

\subsection{WD\,2138$-$332}\label{Sect_WD2138}
\begin{figure}
\includegraphics*[width=9.3cm,trim=0.75cm 6.5cm 0cm 3cm,clip]{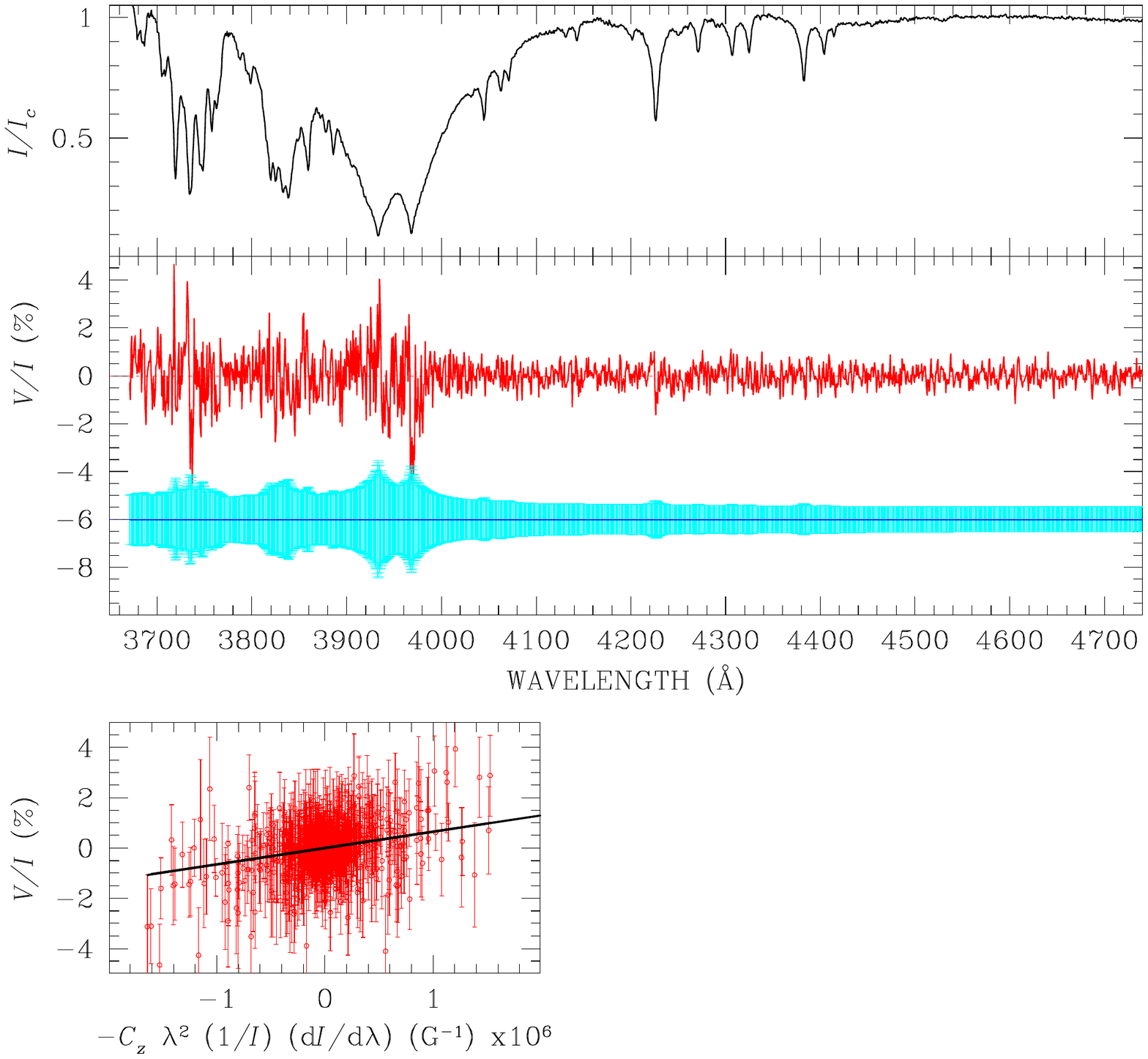}\\
\vskip 0.5 cm

\includegraphics*[width=9.3cm,trim=0.75cm 6.5cm 0cm 3cm,clip]{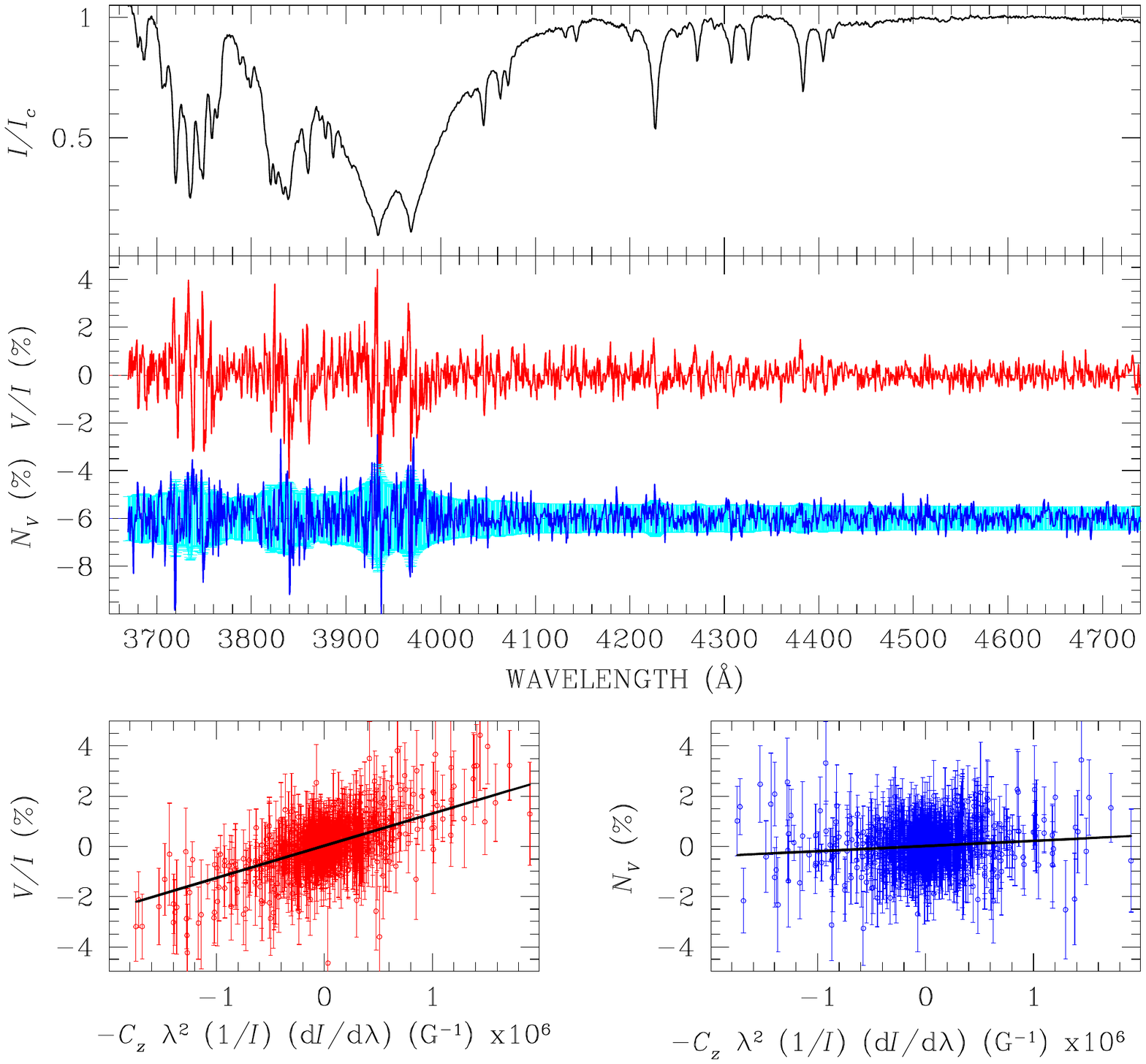}
\caption{\label{Fig_WD2138} 
  Polarised spectra of WD\,2138$-$332 obtained with the FORS2 instrument at two different
  epochs. The upper spectrum, obtained with grism 1200B, was already published
  by \citet{BagLan18} and is reproduced here (with a revised \bz\ measurement) for comparison with the new FORS2 observations obtained in 2018.
}
\end{figure}
WD\,2138$-$332 = L\,570-26 is the hottest star of our sample. Figure~\ref{Fig_WD2138} shows the strong Ca\,{\sc ii} absorption lines at 3933 and 3968\,\AA\ that are clearly evident in the spectrum. A weaker Ca\,{\sc i} line is seen at 4226\,\AA.

\citet{BagLan18} reported a single field measurement of this star, yielding a field detection at the $~\sim 3\,\sigma$ level.  This single detection was not sufficient for us to consider that the star definitely has a weak magnetic field, but the result called for further observations. A single blue spectrum of $I$ and $V/I$ was obtained using FORS2 in October 2018. This new spectrum shows clear and very large circular polarisation signals at the Ca\,{\sc ii} H \& K lines. Nevertheless, as is the case for WD\,1009$-$184, the field is not strong enough to produce clear Zeeman splitting of any of the spectral lines in the $I$ spectrum (Fig.~\ref{Fig_WD2138}). Because the field is so weak, the value of \bs\ cannot be measured, or even detected by examination of the unpolarised $I$ spectrum. The presence of a field is established entirely through the circular polarisation features that coincide with absorption lines. Because the weak-field approximation may be used, we have evaluated \bz\ through Eq.~(\ref{Eq_Bz}), finding $\bz \approx 12.6 \pm 0.6$\,kG. This detection is obtained at a significance of more than $20\,\sigma$, and taken together with the earlier measurement reported by \citet{BagLan18}, clearly establishes the presence of one of the weakest fields known in a WD. 

We note that estimates of the magnetic field from the data obtained in 2014 strongly depend on the decision whether the deep Ca\,{\sc ii} H+K absorption lines are included; in \citet{BagLan18} we used only the spectral region at $\lambda \ga 4110$\,\AA, and reported $\bz = 3.3\,\pm 1$\,kG. However, we realised that the field measured from 3650 to 4150\,\AA\ is three times larger: $\bz = 9.0 \pm 1$\,kG. By combining spectral lines in all regions observed by FORS2, we obtain $\bz = 6.5 \pm 0.7$\,kG. This inconsistency is disappointing, and we do not have a good explanation for it; the 2014 observations were obtained using only two positions of the retarder waveplate, therefore we cannot use the null profiles as a diagnostic tool to verify the presence of a spurious signal. However, because the field is also firmly detected in the observations obtained in October 2018, there is no doubt that the star is magnetic. In \citet{BagLan18} we reported that low \snr\ spectra of this star obtained with the FEROS low-resolution spectrograph were retrieved from the ESO Archive, and that from their analysis we had inferred an upper limit on \bz\ of the order of 20\,kG. We subsequently realised that we had analysed an incorrect spectrum; therefore, our previous estimate of the upper limit for \bs\ is totally incorrect.

WD\,2138--332 has a rich spectrum of Fe~{\sc i} lines around the Ca\,{\sc i} line at 4226\,\AA. These lines show no obvious signs of Zeeman splitting, and so we are only able to provide an upper limit to \bs\ of about 400\,kG, a somewhat larger upper limit than for the other magnetic DZ stars because we do not have any higher resolution ISIS spectra.  

\subsection{New observations of other four DZ and DZA WDs in the local 20\,pc volume}\label{Sect_Others}
We also observed the DZ stars WD\,0840$-$136 with FORS2 and WD\,1705$+$030 with ISIS, and the DZA stars WD\,0738$-$172 and WD\,1626$+$368 with ISIS one time each. The full details of our measurements will be reported in a forthcoming paper, but here we can anticipate that none of them led to a field detection. However, for reasons that will be discussed in Sect.~\ref{Sect_Discussion}, we cannot rule out that in these stars, weak magnetic fields, similar in strength to those found in the stars of Sects~\ref{Sect_WD0816}--\ref{Sect_WD2138}, are present.

\section{Discussion}\label{Sect_Discussion}

The WDs of the 20\,pc volume belong to two rather different groups. The stars whose atmospheres are dominated by hydrogen are clearly identified (at least for values of \teff\ above about 5\,000\,K) by the presence of the Balmer line spectrum. These stars make up 65--70\,\% of the well-identified WDs in the 20\,pc volume, and in general, we can detect magnetic fields down to the kiloGauss level using spectropolarimetry of the Balmer lines. The remaining 30--35\% of  WDs have He-dominated atmospheres. These stars comprise a spectroscopically diverse group that includes all the warmer DC stars, the DQs, the rare DX WDs, and the DZ stars. Magnetic fields in stars of this family can mostly be detected through continuum circular polarisation alone, a technique that is only effective at detecting fields of the order of 10\,MG or more. The one exception to this discouraging limitation on  field detection in He-dominated WDs is provided by the roughly 25\% of these stars that show DZ spectra. In these stars, as we have shown above, fields of a few kiloGauss can be detected. The DZ stars provide the only window into weak magnetic fields that may be present in the much larger family of He-dominated WDs.

\subsection{DZ and DZA stars in the local 20\,pc volume}
We have firmly identified new magnetic fields in four DZ WDs in the local 20\,pc volume.  We note that WD\,1009$-$184 is in a (non-interacting) wide binary system, while the remaining three are single stars \citep[][and references therein]{Tooetal17}.

In addition to these four stars, we are aware of another seven DZ or DZA stars tha are located within a 20\,pc distance from the Sun (see Table~\ref{Table_Other_Stars}), six of which have a He-dominated atmosphere.  None of them has a magnetic field that would be strong enough to be detectable with low- or medium-resolution spectroscopy. However, they may still host a magnetic field with strength comparable to those of the stars of Table~\ref{Table_DZ}. Below we briefly comment on each of them. 

In spite of our null detection with ISIS mentioned in Sect.~\ref{Sect_Others}, WD\,0738$-$172 should still be considered a good candidate (weakly) magnetic DZ star in the 20\,pc volume to be added to the four magnetic stars of Table~\ref{Table_DZ}: on the basis of FORS1 measurements around H$\alpha$, \citet{Frietal04} found that the star is magnetic, although at a low significance level ($\bz = -6.9 \pm 2.1$\,kG). Clearly, this star should be re-observed with high \snr\ spectropolarimetry.

WD\,0046$+$051 = vMa2 is the prototype of metal-enriched WDs. Its strong Ca\,{\sc ii} H+K lines were observed more than one century ago by \citet{van17} \citep[see Fig.~2 of][]{Farihi16}. With ISIS in spectropolarimetric mode, \citet{BagLan18} measured $\bz = -1.7 \pm 1.0$\,kG. Although this is a null detection, the magnetic field could be variable and detectable at other rotation phases, and the star should certainly be re-observed. 
\input{Table_Other_Stars}

The situation is similar for stars WD\,0840$-$136, WD\,1626$+$368, and  WD\,1705$+$030: our FORS2 and ISIS measurements mentioned in Sect.~\ref{Sect_Others} did not lead to a field detection, but these stars should be re-observed once or twice more to sample their rotation cycle, and with higher precision, in particular WD\,1705$+$030, for which our ISIS measurement of Sect.~\ref{Sect_Others} has an uncertainty as large as 5\,kG. In conclusion, for these three stars, a weak field could also have escaped our detection attempts.

To the best of our knowledge, neither WD\,0552$-$041 nor WD\,2251$-070$ have ever been observed in spectropolarimetric mode; from the available low-resolution intensity spectra we can only set an upper limit to its field strength (if a magnetic field exists) of about 0.5\,MG. We note that according to Table~2 of \citet{Giaetal12} or Table~4 of \citet{Subetal17}, WD\,0552$-$041 is the only known H-rich DZ star in the local 20\,pc volume.

\citet{Holletal17} have identified several new WDs within the 20\,pc volume that do not yet have spectral classifications. One or more of these might be DZ stars. Classification and spectro-polarimetric observations of these stars are required. The sample of DZ/DZA stars within 20\,pc of the Sun may not yet be quite complete, but we find that at least 40\,\% of the known WDs with a He-dominated atmosphere with metal lines are magnetic. 

\subsection{High frequency of magnetic fields in DZ and DZA stars }
All DZ WDs that were identified as magnetic in previous work \citep[26 in total, see Table~4 of][]{Holletal17} were discovered by observation of the splitting of their spectral lines, and they typically possess a magnetic field of the order of 0.5 to 10\,MG, while the magnetic field strength of the DZ MWDs discovered in this paper is two orders of magnitude lower.  High-resolution spectroscopy, even when obtained with very high \snr, may detect a magnetic field only when its disc-averaged strength is $\ga 50$\,kG in DA stars, or $\ga 200$\,kG in DZ stars; with SDSS low-resolution spectroscopy, this limit is of the order of ten times higher. Spectropolarimetry carried out with the current instrumentation may detect fields with a disc-averaged line-of-sight component of a few kiloGauss, which may be caused by a dipolar field with a strength at the pole of the order of ten or tens of kiloGauss.  It is therefore natural to expect that spectropolarimetric measurements, sensitive to much weaker magnetic fields than normal spectroscopy, point to an incidence rate for magnetic fields that is substantially higher than the previous estimate of $\sim 10$\,\%, which was based on the analysis of unpolarised profiles. 

In conclusion, with an increased sensitivity of field measurements, we have provided rather strong evidence that the incidence of the magnetic fields in DZ stars is substantially higher than previously estimated.

\subsection{Is the high frequency of the magnetic field linked to spectral pollution or is it an age effect?}
\citet{Kawetal19} proposed that the incidence of magnetic fields in cool DAZ stars is of the order of 50\,\%. The results of our work and those of \citet{Kawetal19} combined suggest that the metal lines and those of the magnetic field are connected to each other, perhaps originating from a common mechanism. \citet{Faretal11} and \citet{Brietal18} both suggested that in cool stars, a weak magnetic field may be generated by the plunging of a giant gaseous planet onto the star. However, we need to emphasize that because of small number statistics, there is still no really compelling evidence that magnetic fields are more frequent in WDs of certain spectral types or temperature ranges. For instance, it might be argued that our statistics on DZ stars is still at least marginally consistent with the statistics of DA WDs reported by \citet{LanBag19b}, who found that $20 \pm 5$\,\% of DA WDs in the local 20\,pc volume are magnetic. Even assuming a scenario in which magnetic fields are more common in DZ than in DA stars, which needs to be supported by further data, it is not obvious, however, that a magnetic field is linked to accreted metal pollution in an He-dominated atmosphere. Although it might be tempting to search for a mechanism that is responsible for accretion from planetary or asteroid debris and also for a magnetic field, we recall that without metal lines, the stellar spectrum of a cool WD with an He-dominated atmosphere would appear totally featureless. From a featureless spectrum, a magnetic field may be detected only if it is strong enough to polarise the continuum, therefore a magnetic field weaker than a few tens of megaGauss would escape detection. It is possible instead that the higher incidence of magnetism in DZ stars arises because magnetic fields are more frequent in cooler WDs than in hotter WDs, regardless of whether they have a polluted atmosphere. The magnetic field may only be revealed when metal lines are present in their spectra, however. That temperature and age are important discriminating factors is also suggested by the fact that a magnetic field is much more frequent in DZ stars cooler than $\sim 8\,000$\,K than in DZ stars hotter than $\sim 8\,000$\,K \citep{Holletal15}.

The alternative hypothesis that the frequency of the magnetic field is higher in cooler and older WDs than in hotter and younger WDs has frequently been suggested and discussed in the past \cite[e.g.][]{LiebertSion79,FabVal99,Lieetal03,KawVen14}. The reality is that the interpretation of literature data is still very problematic because different surveys are affected by a number of different biases, such as target selection, observing technique, or low sensitivity. As explained in Sect.~\ref{Sect_Intro}, our volume-limited spectropolarimetric survey is motivated by the necessity to obtain a decent-sized sample of statistically unbiased data. \citet{LanBag19b} have presented some preliminary data obtained from this survey, and proposed that the production of DA MWDs in the local 20\,pc volume has been roughly constant with time. Because younger WDs are more abundant than older WDs \citep[the frequency of creation of new WDs has steadily increased with time, see Sect.~3.3.3 of][]{Holbetal16}, the observational evidence presented by \citet{LanBag19b} leads to the conclusion that the fraction of magnetic stars in the local 20\,pc volume increases with age. This scenario is re-enforced by our discovery of a magnetic field in four DZ WDs belonging to the same volume, the youngest of which is $\sim 2$ Gyr old.

\section{Conclusions}
We have reported the discovery of four weakly magnetic DZ WDs, all within 20\,pc distance from our Sun. To the best of our knowledge, these are the first DZ WDs discovered to be magnetic solely through spectropolarimetry of their metal lines. In this respect, our FORS2 spectrum of WD\,1009$-$184 represents a very significant example of how a relatively weak magnetic field that is barely able to split metal lines if observed in (unpolarised) intensity, may be easily detected with low-resolution circular spectropolarimetry. More generally, it is evident that highly sensitive spectropolarimetric measurements carried out in the past decade or so fundamentally contribute to clarifying the actual occurrence of magnetism in WDs of various spectral types and ages. This insight can hardly be obtained by using spectroscopic data alone because too many WDs appear to possess weak magnetic fields that are not strong enough to show Zeeman split lines in (unpolarised) spectroscopy.

We have found that at least 40\,\% of DZ stars with an He-rich atmosphere within 20\,pc from the Sun are magnetic, and we suggested that further observations could establish that the fraction of magnetic DZ WDs in the local volume is in fact even higher. Our small number statistics leaves the overall incidence of magnetic fields in DZ WDs only roughly defined, but we note that the fact that DZ WDs have a higher incidence of magnetic fields than WDs of other spectral type is also supported by the finding of \citet{Holletal15} and \citet{Holletal17} that about 10\,\% of DZ stars have a spectrum with Zeeman split lines. We argue that \citet{Holletal15} and \citet{Holletal17} have detected just the fraction of magnetic DZ stars in the regime between 1 and 100\,MG, but that a large fraction of DZ stars exhibit a magnetic field that may be revealed only through circular spectropolarimetry. That the presence of a magnetic field in a DZ star is more the rule than the exception is a hypothesis that deserves to be investigated with further high-sensitivity spectropolarimetric observations of DZ WDs outside of the 20\,pc volume. 

We cannot establish whether a magnetic field is directly connected to the accretion mechanism that explains the presence of metal lines, or if it is simply the presence of metal lines that allows us to detect magnetic fields that are more frequent in cool WDs than in other types of WDs. However, because \citet{LanBag19b} have found that the production rate of magnetic DA WDs seems constant with time, and because young WDs are more frequent than old ones, we deduce that the frequency of magnetic fields is higher in older than younger WDs. Having secured enough telescope time to complete our 20\,pc volume limited survey, we will be able to present more quantitative conclusions in the near future.

\begin{acknowledgements}
This work is based on observations made with ESO Telescopes at the
La Silla Paranal Observatory under programme IDs 0102.D-0045 and 0103.D-0029,
on data obtained from the ESO Science Archive Facility,
and on observations collected at the William Herschel
Telescope, operated on the island of La Palma by the Isaac Newton
Group, programme P15, during semester 18B and programme P10 during
semester 19A. JDL acknowledges the financial support of the Natural Sciences
and Engineering Research Council of Canada (NSERC), funding reference
number 6377-2016.
\end{acknowledgements}


\bibliography{sbabib}

\end{document}

%% file: Table_Stars.tex
\begin{table*}
\begin{center}
\caption{\label{Table_DZ} Parameters of the known magnetic DZ stars within 20\,pc from the Sun.}
\begin{tabular}{lrrlrrrrrrl}
\hline\hline
               &
               &
$d$            &
               &
$T_{\rm eff}$  &
               &
               &
Age            &
               &
               \\
STAR           &
$V$            &
(pc)           &
Spectrum       &
(K)            &
$\log g$       &
$M/M_\odot$    &
(Gyr)          &
$\log(L/L_\odot)$ &
Reference       \\
\hline
WD\,0816$-$310 &15.4 &19.36& DZ7.8 & 6463 & 8.00 & 0.57 & 2.02 &$ -3.61 $& 2,3 \\
WD\,1009$-$184 &15.4 &18.09& DZ8.3 & 6036 & 8.02 & 0.59 & 2.63 &$ -3.74 $& 2,3 \\
WD\,1532$+$129 &15.7 &19.27& DZ9.3 & 5430 & 7.90 & 0.51 & 3.36 &$ -3.86 $& 1,3 \\
WD\,2138$-$332 &14.5 &16.11& DZ6.8 & 7399 & 8.19 & 0.70 & 1.87 &$ -3.48 $& 2,3 \\
\hline
\end{tabular}
\tablefoot{Key to references: 
1: \citet{Subetal17};\  
2: \citet{Giaetal12};\
3: Gaia Collaboration (\citeyear{gaia18}).} 
\end{center}
\end{table*}

%% file: Table_Log.tex
\begin{table*}
\caption{New magnetic measurements of DZ WDs.}
\label{Tab_Log}
\centering
\begin{tabular}{llllccrr@{$\pm$}l r@{$\pm $}l r}
\hline\hline
Star          & Instrument & Grism/     &             &\multicolumn{1}{c}{Date} & UT & Exp.&
\multicolumn{2}{c}{\bz}&\multicolumn{2}{c}{\nnz}\\
&            & Grating   &   MJD        &  yyyy-mm-dd & hh:mm&
\multicolumn{1}{c}{(s)}&
\multicolumn{2}{c}{(kG)}&
\multicolumn{2}{c}{(kG)}\\
\hline
WD\,0816$-$310 & FORS1 &600B       & 54395.355 & 2007-10-22& 08:40 &  960 &$-20.8$&2.0&\multicolumn{2}{c}{}\\
               & ISIS &R600B+R1200R& 58592.887 & 2019-04-19& 21:16 & 3600 &$ 18.5$&5.1&$ 6.5$&4.6\\[2mm]
WD\,1009$-$184 & FORS2& 1200B      & 58477.297 & 2018-12-25& 07:07 & 3600 &$-47.4$&1.4&$ 2.0$&1.4\\
               & ISIS &R600B+R1200R& 58592.969 & 2019-04-19& 23:15 & 3600 &$-16.1$&4.6&$-4.0$&4.5\\[2mm]
WD\,1532$+$129 & FORS2& 1200B      & 58560.772 & 2019-03-18& 06:30 & 2700 &$ -4.3$&1.0&$-1.2$&0.9\\
               & ISIS &R600B+R1200R& 58593.106 & 2019-04-19& 02:33 & 4800 &$ -8.5$&1.9&$-0.2$&1.8\\
               & FORS2& 1200B      & 58653.102 & 2019-06-19& 02:27 & 2640 &$-21.5$&0.9&$-0.2$&0.9\\[2mm]            
WD\,2138$-$332 & FORS2& 1200B      & 56812.285 & 2014-06-04& 06:51 & 2400 &$ +6.5$&0.7\\
               & FORS2& 1200B      & 58397.135 & 2018-10-06& 03:14 & 2700 &$+12.8$&0.6&$ 2.0 $&0.6\\
\hline
\end{tabular}
\end{table*}

%% file: Table_Other_Stars.tex
\begin{table*}
\caption{\label{Table_Other_Stars} Known DZ and DZA WDs within 20\,pc from the Sun for which we need
more highly-sensitive spectropolarimetric measurements. Key to references given in Table~\ref{Table_DZ}.}
\centering
\begin{tabular}{lrrlrrrrrl}
\hline\hline
               &
               &
$d$            &
               &
$T_{\rm eff}$  &
               &
               &
Age            &
               &
               \\
STAR           &
$V$            &
(pc)           &
Spectrum       &
(K)            &
$\log g$       &
$M/M_\odot$    &
(Gyr)          &
$\log(L/L_\odot)$ &
Reference       \\
\hline
WD\,0046$+$051 &12.4& 4.32& DZ8.2 & 6130 & 8.16 & 0.67 & 3.45 &$ -3.79 $& 1,3  \\ 
WD\,0552$-$041 &14.5& 6.44& DZ9.3 & 5430 & 8.49 & 0.90 & 7.34 &$ -4.20 $& 1,3  \\ 
WD\,0738$-$172 &13.1& 9.16& DZ6.5 & 7700 & 8.05 & 0.61 & 1.31 &$ -3.33 $& 1,3  \\
WD\,0840$-$136 &15.7&14.80& DZ10.1& 4980 & 7.95 & 0.54 & 5.28 &$ -4.04 $& 1,3  \\ 
WD\,1626$+$368 &14.0&15.89& DZA5.9& 8507 & 8.00 & 0.58 & 1.02 &$ -3.13 $& 2,3  \\ 
WD\,1705$+$030 &15.2&17.87& DZ7.7 & 6584 & 8.18 & 0.68 & 2.75 &$ -3.67 $& 2,3  \\ 
WD\,2251$-070$ &15.7& 8.53& DZ12.6& 4000 & 7.92 & 0.52 & 7.13 &$ -4.40 $& 1,3  \\ 
\hline
\end{tabular}
\end{table*}